\documentclass[twocolumn,
amsmath, 
amssymb, 
aps, 
nofootinbib,
raggedbottom,
superscriptaddress]{revtex4-2}
\usepackage{graphicx} 
\usepackage{preamble}
\usepackage{color}
\usepackage[dvipsnames]{xcolor}
\usepackage[hidelinks]{hyperref}
\usepackage{cancel}
\DeclareMathOperator{\GCD}{gcd}
\newcommand{\avg}[1]{\left\langle#1\right\rangle}
\DeclareMathOperator{\Link}{Link}

\newcommand{\diff}{\mathrm{d}}

\begin{document}
 
 \title{\boldmath{\texorpdfstring{Phases of theories with $\Z_{{} N}$ 1-form symmetry,  \\ and   the  roles of center vortices and magnetic monopoles}{The phases of theories with $\Z_{{} N}$ 1-form symmetry, \\ and the roles of center vortices and magnetic monopoles}}}
\author{Mendel Nguyen}
\affiliation{Department of Physics, North Carolina State University, Raleigh, NC 27607, USA \looseness=-1}
\author{Tin Sulejmanpasic}
\affiliation{Department of Mathematical Sciences, Durham University, Durham DH1 3LP, UK \looseness=-1}
\author{Mithat \"{U}nsal}
\affiliation{Department of Physics, North Carolina State University, Raleigh, NC 27607, USA \looseness=-1}

\begin{abstract}
We analyze the phases of theories which only have a  microscopic $\Z_{{} N}$ 1-form symmetry, starting with a topological \emph{BF}  theory and deforming it in accordance with microscopic symmetry. These theories  have a well-defined notion of confinement. Prototypical examples are pure $\SU(N)$ gauge theories and $\ZZ_{{} N}$ lattice gauge theories. 
Our analysis shows that the only generic phases are in $d=2$, only the confined phase; in $d=3$, both the confined phase and the topological \emph{BF} phase; and in $d=4$, the confined phase, the topological \emph{BF} phase, and a phase with a massless photon. 
We construct a $\Z_{{} N}$ lattice gauge theory  with a deformation which, surprisingly, produces up to $(N-1)$ photons. We give an interpretation of these findings in terms of two competing pictures of confinement -- proliferation of  monopoles and proliferation of  center vortices -- and conclude that the proliferation of center vortices is a necessary but insufficient condition for confinement, while that of monopoles is both necessary and sufficient. 
\end{abstract}

\maketitle
\newpage

\section{Introduction}
To this day, a completely general understanding of quark confinement remains elusive.
Much work over the past several decades has been devoted to unraveling the underlying mechanisms at play. 
One of the key ideas has been the notion of 1-form symmetry, which allows us to characterize confinement as a phase where such a symmetry is unbroken \cite{tHooft:1977nqb,Polyakov:1978vu,Gaiotto:2014kfa}. 
For $\SU(N)$ gauge theory with adjoint matter, this 1-form symmetry is $\Z_{{} N}$, and historically has gone by the name `center symmetry.'
While QCD itself does not have such a symmetry, being explicitly broken by fundamental matter,\footnote{
This is not to say that confinement has no meaning in theories with fundamental matter such as QCD -- a perspective which is sometimes conflated with the absence of 1-form symmetry. Confinement in QCD is the statement that there are no asymptotic quark states.
}
it is safe to say that understanding systems with $\ZZ_{{} N}$ 1-form symmetry should go a long way towards understanding confinement. 
In this work we  focus exclusively on such systems.
In the '70s and '80s, the predominant picture of quark confinement was the so-called dual-superconductor picture in which magnetic monopoles condense, causing the expulsion of electric field via the dual Meissner effect \cite{Mandelstam:1974pi, Polyakov:1975rs, tHooft:1977nqb, tHooft:1979rtg, tHooft:1981bkw}. 
Indeed, this picture is reinforced by the fact that in the Higgs regime of $\SU(N)$ gauge theory with adjoint scalar matter, where the low energy dynamics is that of an Abelian gauge theory, there exist magnetic monopole solutions. 
In 3d, monopoles are instanton-like objects (i.e., points in space-time) and Polyakov \cite{Polyakov:1975rs, Polyakov:1976fu} showed that they are generically relevant, leading to mass gap and confinement.
A similar thing happens in 4d if the monopoles condense \cite{Seiberg:1994rs, Unsal:2008ch, Unsal:2007jx}. 
The theory can then be cast in electric--magnetic dual language, with monopoles described as elementary scalars. 
Static electric charges are then connected via Abrikosov--Nielsen--Olesen strings, and confinement is manifest.

There is, however, a competing picture of the confinement mechanism in which the primary agents are not monopoles but center vortices \cite{ tHooft:1977nqb,Cornwall:1979hz, Nielsen:1979xu}.
These are objects whose worldvolumes are codimension-2 in spacetime, and their linking with Wilson loops are accompanied by a $\Z_{{} N}$ phase. 
Roughly, the idea behind the center vortex mechanism is analogous to the mechanism by which discrete 0-form symmetries are restored. 
When a discrete 0-form symmetry is spontaneously broken, there exist heavy domain walls, and the restoration of the symmetry can be seen when these domain walls become light and proliferate.
Center vortices are meant to be for center symmetry what domain walls are for discrete 0-form symmetry -- 
they are heavy when the center symmetry is spontaneously broken, and their proliferation ostensibly signals the restoration of the center and hence the onset of confinement. 

Unlike the monopole mechanism, however,
there exists no analytical treatment of the center-vortex scenario except in $d=2$ \cite{Callan:1977qs,Tanizaki:2022ngt}. 
Our understanding of center vortex effects is based primarily on heuristic arguments and numerical studies (see \cite{DelDebbio:1996lih,DelDebbio:1998luz,Engelhardt:1998wu,Alexandrou:1999iy,Alexandrou:1999vx,deForcrand:1999our,deForcrand:2000pg,Sale:2022qfn} for an incomplete list of references).
It is therefore desirable to find new methods that allow us to determine the roles of vortices and their relation to monopoles in a reliable analytic fashion. 
   
In this work, we propose one such method, which allows us to understand the precise roles of both center vortices and magnetic monopoles in determining the phases of systems with $\Z_{{} N}$ 1-form symmetry.
To do this, it is necessary to consider a framework in which these objects coexist. 
Such a framework arises naturally by developing an effective field theory for the $\Z_{{} N}$ 1-form symmetry. 
Our starting point is the assumption that the system can be taken to a regime where the system is gapped and the $\Z_{{} N}$ 1-form symmetry is spontaneously broken.\footnote{For instance, pure $\SU(N)$ gauge theory can be deformed by introducing two adjoint scalars which may Higgs $\SU(N)$ to $\Z_{{} N}$.}
In such a regime, the correct deep infrared description is provided by the topological \emph{BF} theory, which will be reviewed below. 
However, the \emph{BF} theory has more symmetry than the $\Z_{{} N}$ 1-form symmetry assumed microscopically.
For this reason, we must deform the \emph{BF} theory in such a way that only the assumed microscopic symmetry is present. 
As we will argue below, these deformations may be interpreted as the coupling to center vortices and magnetic monopoles. 
From the resulting deformed theory, we can then determine the minimal number of generic phases the system should have, as in the idea of the deformation class \cite{Seiberg_DC}.
For $d=4$, we find that the generic phases are characterized by the following behaviors:
\begin{enumerate}
\item[(a)] Neither vortices nor monopoles proliferate.
\item[(b)] Both vortices and monopoles proliferate.
\item[(c)] Vortices proliferate and monopoles do not.
\end{enumerate}
Notice that there is no phase where monopoles proliferate and vortices do not. 
The reason for this is that monopoles are attached to $N$ center vortices and are always heavy when center vortices are heavy. 
The phase (a) is just the deconfined, topological phase. 
The phase (b) is the confining phase, as both monopoles and center vortices proliferate. 
But, as we will see, the phase (c) is the photon phase, which conclusively shows that, while the proliferation of center vortices is a necessary condition for confinement, it is not sufficient.

\section{Deformations of \emph{BF} theories}

In order to make our points transparent, we list our assumptions here:
\begin{itemize}
\item We assume the existence of $\Z_{{} N}$ 1-form and time-reversal symmetry, without any mixed or pure 't Hooft anomalies,
\item We assume that Poincare symmetries are either exact or emerge in the IR, and are not spontaneously broken. 
\end{itemize}

The \emph{BF} model that describes the gapped phase of broken $\Z_{{} N}^{[1]}$ symmetry\footnote{We write $G^{[p]}$ for a $p$-form symmetry group $G$.} is given by the (Euclidean) Lagrangian
\begin{equation}\label{eq:S_BF}
\mathcal{L}_{\emph{BF}} = \frac{iN}{2\pi} \, b \, \diff a
\end{equation}
where $a$ and $b$ are $1$-form and $(d-2)$-form gauge fields, $d$ being the spacetime dimensionality.
This model has the following symmetries:
\begin{alignat}{3}
    &\Z_{{} N}^{[1]}  &&\colon a \to a + \alpha \qquad &&(\diff \alpha = 0, \   \textstyle{\oint} \alpha \in \Z_{{} N}) \\
    &\hat{\Z}_{{} N}^{[d-2]} &&\colon b \to b + \beta \qquad &&(\diff\beta = 0, \   \textstyle{\oint} \beta \in \Z_{{} N})
\end{alignat}
The $\hat{\Z}_{{} N}^{[d-2]}$ symmetry, however, is not assumed microscopically. 
In order to explicitly break the $\hat{\Z}_{{} N}^{[d-2]}$ symmetry, we introduce objects that are charged under the gauge group of $b$.
These objects necessarily have worldvolumes $V$ of codimension 2, and thus introduce into the path integrand a factor $\exp(i\int_V b)$. 
Since the correlation function $\langle \exp(i\int_V b) \exp(i\int_C a) \rangle$ in the \emph{BF} theory gives $\exp \{i \frac{2\pi}{N} \Link(V,C)\}$ (see Appendix \ref{sec:BF_review}), we naturally interpret these objects as center vortices, as this is the defining property of such objects. 
Note that invariance under gauge transformations $b \to b+\diff\omega$ requires that the center-vortex worldvolume $V$ be closed. 

\begin{figure}[tbp] 
   \centering
   \includegraphics[width=3.5in]{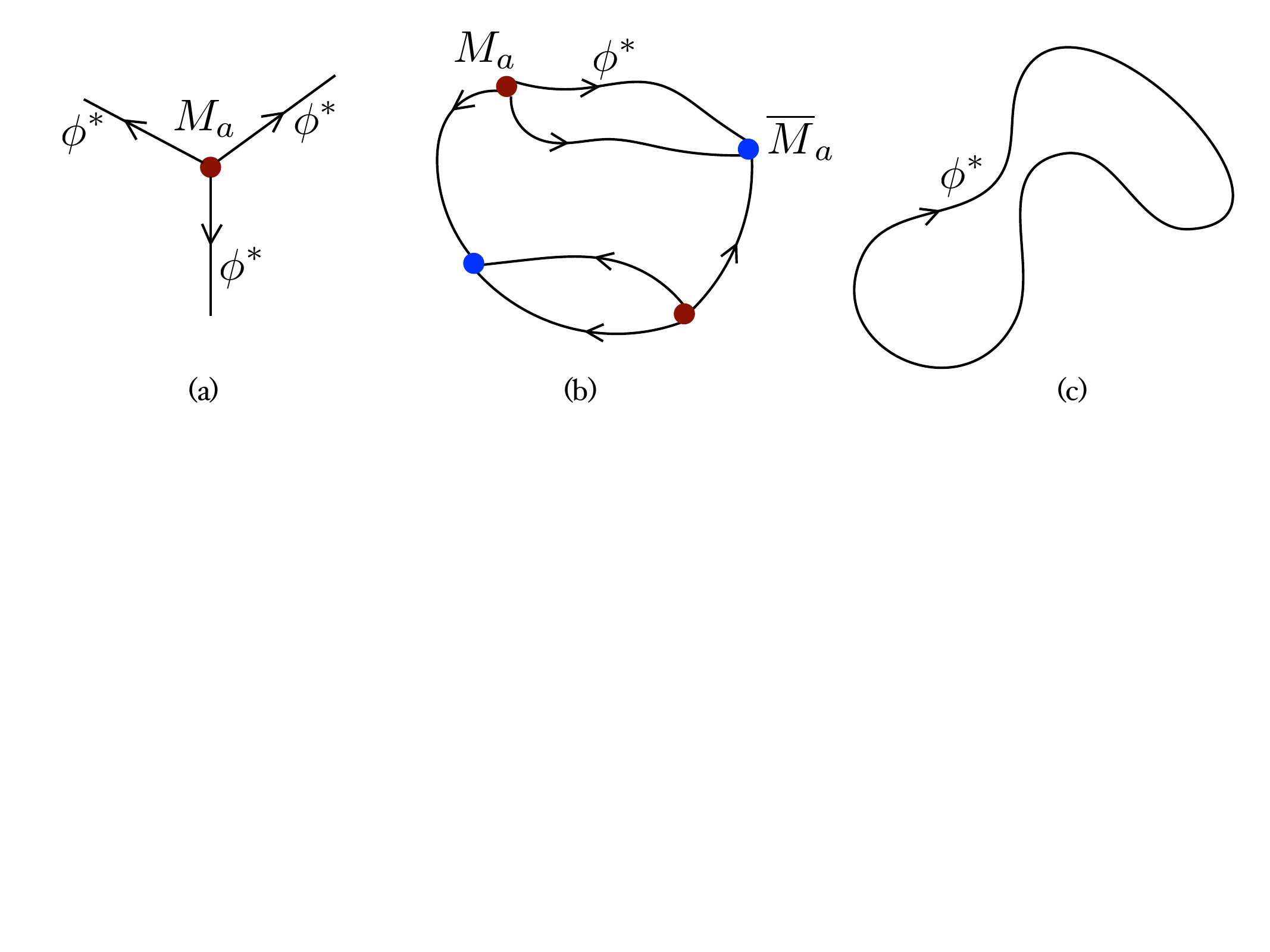} 
   \vspace{-4cm}
   \caption{a) A monopole sources $N$ vortices. (We set  $N=3$ above).     b) A generic configuration of monopoles and vortices.
   c) Vortices can exist independent of monopoles. Diagrams correspond to either 3d Euclidean space-time or spatial slice in a 4d theory.  }
   \label{fig:vm}
\end{figure}

Having introduced center vortices, however, we actually gain an additional `magnetic' $\tilde{\U}(1)^{[d-3]}$ symmetry associated with the closedness of $\diff a$ when $d\geq3$. 
Note that before we introduced the center vortices which couple to $b$, this magnetic symmetry was not present. 
This is because the would-be operators charged under this symmetry, the monopoles of $a$, are charged under the gauge group of $b$. 
To see this, consider a monopole configuration, $\diff(\diff a)=2\pi \delta[M]$, where $\delta[M]$ is a 3-form supported on the monopole worldvolume $M$.
The monopole configuration is clearly not invariant under the gauge transformation $b\rightarrow b+\diff\omega$, as it causes the \emph{BF} action to transform as $S_{\emph{BF}} \to S_{\emph{BF}} + i N \int_M S_M \omega$. 
This can be fixed however by allowing open center vortex configurations with $\del V = NM$. 
It follows that in the presence of dynamical vortices, one can define a monopole operator by attaching it to the junction of  $N$ center vortices [see Fig.\ref{fig:vm}(a)],\footnote{This observations is not new and has been made in many works including \cite{tHooft:1977nqb,Yoneya:1978dt,DelDebbio:1997ke,Alexandrou:1999iy,Greensite:2003bk,Deldar:2015kga}.}
and this is the reason that the magnetic $\tilde{\U}(1)^{[d-3]}$ symmetry becomes a genuine global symmetry once center vortices are introduced.
But again, under our assumption, such a symmetry is not microscopic, and  we must explicitly break this additional symmetry, by the introduction of dynamical monopoles into the system. 
Note that since monopoles are each attached to $N$ vortices, monopoles are heavy whenever vortices heavy.
Therefore, there is no phase where monopoles proliferate but vortices do not.
This explains the three phases (a), (b) and (c) we discussed in the introduction.

Let us now consider the cases $d=2,3,4$ separately to discuss the various phases that can arise. 

\subsection{Two dimensions}

In 2d, center vortices are instantons and monopoles do not exist.
The \emph{BF} theory has $\Z_{{} N}^{[1]} \times \hat{\Z}_{{} N}^{[0]}$ symmetry.
The coupling to center vortices is accomplished by simply adding the operators $\exp(\pm ib)$ to the \emph{BF} action \eqref{eq:S_BF}.
Indeed, this explicitly breaks the $\hat{\Z}_{{} N}^{[0]}$ symmetry under the shift $b\to b+\frac{2\pi}{N}$.
Moreover, this perturbation is relevant, as the operators $\exp(\pm ib)$ are topological in the \emph{BF} theory, and hence have vanishing scaling dimension \cite{Cherman:2021nox}.
The resulting effective Lagrangian is therefore
\begin{align}
    \mathcal{L} = \frac{iN}{2\pi}\,b\,\diff a - 2 \zeta \cos b
\end{align}
It is now easy to see that the assumption that the system can be pushed to a regime of spontaneously broken $\Z_{{} N}^{[1]}$ is not robust.
Indeed, as pointed out in Ref.~\cite{Cherman:2021nox}, in 2d the perturbation by the generators of the $\Z_{{} N}^{[1]}$ symmetry will generically lead to $\Z_{{} N}^{[1]}$ restoration, i.e., confinement.
This can be seen by explicitly computing Wilson loop averages and observing the area law (see Appendix \ref{sec:2dconf}).

Alternatively, we can work in the Hamiltonian framework (see Appendix \ref{sec:BF_review}). 
On a spatial circle, the Hilbert space is spanned by the eigenstates $|q\rangle$ of $\exp(ib)$ with eigenvalues $\exp(i \frac{2\pi}{N}q)$, and the action of $\exp(i \oint a)$ on these states shifts $q$ by $1$.
These states are degenerate in the \emph{BF} theory, but become nondegenerate under the perturbation, as the Hamiltonian density becomes $-2 \zeta \cos b$. 
In particular, the unique ground state is given by $|q=0\rangle$, which is invariant under the $\Z_{{} N}^{[1]}$ symmetry operator $\exp(ib)$.
This means that $\Z_{{} N}^{[1]}$ is unbroken.  

\subsection{Three dimensions}

In 3d, vortices are particles and monopoles are instantons.
The \emph{BF} theory now has $\Z_{{} N}^{[1]} \times \hat{\Z}_{{} N}^{[1]}$ symmetry.
Because we do not assume the $\hat{\Z}_{{} N}^{[1]}$ symmetry microscopically, let us break it explicitly by introducing a complex scalar $\phi$ with unit charge under the gauge group of $b$, and add to the Lagrangian a term\footnote{Note that our assumption of $T$-symmetry prohibits Chern--Simons terms.}
\begin{align}
    \mathcal{L}_1 = \frac{1}{2 g^2} |\diff b|^2 + |(\diff - i b) \phi|^2 + \ldots 
\end{align}
where we have also included a kinetic term for $b$.\footnote{
Kinetic terms for all fields will always be generated. 
While they are irrelevant with respect to the \emph{BF} term, being higher in derivatives, we include them because they can be \emph{dangerously} irrelevant.
} 
Having done this, however, we now gain a magnetic $\tilde{\U}(1)$ 0-form symmetry associated with the closedness of $\diff a$.
This is a phenomenon that we alluded to in start of this section, but did not encounter in the 2d case, so let us take a moment to explain it here in detail.
Let $M_a$ denote the operator for unit monopoles of $a$.
The effect of inserting $M_a(x_0)$ under the path integral is to impose the condition $\diff (\diff a)(x) = 2 \pi \delta^3(x-x_0)$ for $x$ near $x_0$. 
Under a gauge transformation $b \to b + \diff \omega$, the change in the \emph{BF} action in the presence of the $M_a(x_0)$ insertion is given by
\begin{align}
    \frac{i N}{2\pi} \int \diff \omega \, \diff a= - \frac{i N}{2\pi} \int \omega \, \diff(\diff a) = - i N \omega(x_0)
\end{align}
This shows that $M_a$ carries gauge charge $N$ coupled to $b$.
Therefore, in the theory defined by the \emph{BF} Lagrangian alone, there can be no magnetic $\tilde{\U}(1)$ 0-form symmetry, as the only potentially charged operator, $M_a$, is not gauge invariant. 
With the inclusion of the scalar field $\phi$, however, we can dress the monopole operator to form the gauge-invariant operator $M_a (\phi^*)^N$, and this gives rise to the additional $\tilde{\U}(1)$ magnetic symmetry. 
As we do not assume this $\tilde{\U}(1)$ symmetry microscopically, let us therefore add to Lagrangian
\begin{align}
    \mathcal{L}_2 = \frac{1}{2e^2}|\diff a|^2 + \zeta  M_a (\phi^*)^N + \cc 
\end{align}
where we have also included a kinetic term for $a$. 
The deformed theory defined by the Lagrangian $\mathcal{L} = \mathcal{L}_{\emph{BF}} + \mathcal{L}_1 + \mathcal{L}_2$ has only the symmetry we have assumed microscopically, $\Z_{{} N}^{[1]}$. 

Let us now examine the phases of the deformed theory.
If $\phi$ is heavy, then upon integrating it out, 
the deformations $\mathcal{L}_1$ and $\mathcal{L}_2$ both drop out, and we return to the \emph{BF} theory (Higgs phase) we started with. 
If $\phi$ condenses, however, then $b$ becomes heavy, and now the \emph{BF} term drops out instead.
The effective Lagrangian now reads 
\begin{align}
    \mathcal{L} \to \frac{1}{2e^2}|\diff a|^2 + \zeta \langle \phi^* \rangle^N M_a + \cc
\end{align}
which describes a pure $\U(1)$ gauge theory perturbed by monopole operators. 
As Polyakov showed \cite{Polyakov:1975rs,Polyakov:1976fu}, these monopole operators are relevant and lead to a phase of unbroken $\Z_{{} N}^{[1]}$, i.e., confinement.
Note that a Coulomb phase cannot generically arise unless we assume the $\tilde{\U}(1)$ magnetic symmetry microscopically, forbidding the inclusion of monopole operators in the effective Lagrangian. 

\subsection{Four Dimensions}

We now come to the most interesting case of four dimensions. 
Our discussion here will be brief as we will consider 4d $\Z_{{} N}$ lattice gauge theory as a concrete example in the next section.
In 4d, vortices are strings and monopoles are particles, and the \emph{BF} model has $\Z_{{} N}^{[1]} \times \hat{\Z}_{{} N}^{[2]}$ symmetry.
As in $d=2,3$, to break the nonmicroscopic $\hat{\Z}_{{} N}^{[2]}$ symmetry explicitly, we introduce vortices carrying unit gauge charge coupled to $b$.
As in the $d=3$ case, this introduces an unwanted $\tilde{\U}(1)^{[1]}$ magnetic symmetry, which we then break by introducing unit monopoles of $a$ attached to $N$-fold junctions of vortices. 

Let us now examine the phases of the resulting deformed theory.
\begin{enumerate}
\item[(a)] If the vortices coupling to $b$ are heavy, then the monopoles are heavy too, being always connected to vortices. 
So integrating out vortices and monopoles, we are left with the \emph{BF} theory we started with, which is the topologically ordered Higgs phase.  One can also characterize this phase by the emergence of a $\Z_{{} N}^{[2]}$ 2-form symmetry.\footnote{The emergence of 1-form symmetry is a subtle issue, but Ref.~\cite{Cherman:2023xok} points out that it is robust when the would-be emergent symmetry is spontaneously broken or participates in a mixed anomaly, as is the case here and everywhere else we speak of emergent 1-form symmetry.}
\end{enumerate}
Next, consider the case where vortices proliferate. 
Then the 2-form gauge field $b$ acquires a mass by the Higgs mechanism, so that the \emph{BF} term drops out of the Lagrangian. 
We now have two sub-cases to consider.
\begin{enumerate}
\item[(b)] If the monopoles condense, then the system confines, i.e., $\Z_{{} N}^{[1]}$ is unbroken, and the phase has no emergent symmetries and anomalies.
\item[(c)] If the monopoles are heavy, then the magnetic $\tilde{\U}(1)^{[1]}$ symmetry for $a$ emerges in the IR.
At the same time, the fact that the $\emph{BF}$ term is trivial implies that the electric $\Z_{{} N}^{[1]}$ symmetry we started with gets enhanced to $\U(1)^{[1]}$.
The electric and magnetic 1-form symmetries have a mixed anomaly between each other, which can only be satisfied by the existence of a massless photon. 
\end{enumerate}

We note that other scenarios are also possible, such as a phase with multiple flavors of photons, or a partially broken $\Z_{{} N}^{[1]}$ symmetry. 
These are briefly discussed in Appendix~\ref{app:photons}.

\section{4D \boldmath{\texorpdfstring{$\Z_{{} N}$}{ZN}} lattice gauge theory}

One of the more surprising conclusions of our analysis in four dimensions is that the appearance of a Coulomb phase is generic. 
But if this is true, than every system with a $\Z_{{} N}^{[1]}$ symmetry should have a generic Coulomb phase. 
While $\SU(N)$ gauge theories are well known to have such phases, $\Z_{{} N}$ lattice gauge theories are only known to have such phases when $N$ is sufficiently large  \cite{Elitzur:1979uv,Horn:1979fy,Frohlich:1982gf}. 
(In particular, it was shown to be the case numerically when $N\ge 5$ \cite{Creutz:1979zg}.)
Our arguments, however, apply for any value of $N$ (e.g., $N=2$). 
It is the suppression of monopoles together with the proliferation of center vortices that enables a Coulomb phase to emerge. 
We will indeed argue that such a deformation of $\Z_{{} N}$ lattice gauge theory exists, and we will explicitly prove this is the case in the Villain formulation of the lattice gauge theory. 
The argument will consist in establishing a duality with a $\U(1)$ lattice gauge theory, where the existence of a Coulomb phase is manifest.

The Villain formulation of the $\Z_{{} N}$ lattice gauge theory consists of integers $n_l$ on links and integers $m_p$ on plaquettes with the action given by
\begin{align}
    S_{\text{V}} = \frac{\beta}{2} \sum_p |(\diff a)_p + 2 \pi m_p|^2
\end{align}
where $a_l \equiv \frac{2\pi}{N}n_l$.\footnote{We use $l,p,c$ to denote links, plaquettes, and cubes. For variables $v_{\sigma_r}$ on $r$-cells $\sigma_r$, we write $(\diff v)_{\sigma_{r+1}} \equiv \sum_{\sigma_r \in \del \sigma_{r+1}} v_{\sigma_r}$.}
This action can be obtained by taking the large $\beta$ limit of the Wilson action
\begin{align}
    S_{\text{W}} = \beta \sum_p \{1 - \cos (\diff a)_p\}
\end{align}
but we expect both formulations to exhibit the same phase structure.

In either formulation, a single unit center vortex is a configuration where for some closed surface $\Tilde{\Sigma}$ on the dual lattice, we have $(\diff a)_p = \pm \frac{2\pi}{N}$ if $p \in \pm {*\Tilde{\Sigma}}$,\footnote{The dual of an $r$-cell $\sigma_r$ in the original lattice, denoted $*\sigma_r$, is the unique $(d-r)$-cell in the dual lattice that intersects $\sigma_r$ at a point. The dual of a cell in the dual lattice is similarly defined.} and $(\diff a)_p=0$ otherwise. 
But the advantage of the Villain formulation is that the presence of monopoles is manifest.
The magnetic current is simply $\diff m$.
Nevertheless, monopoles can still be identified in the Wilson formulation, as will be described below. 

As we wish to isolate the effect of vortices from monopoles, let us first consider the situation where monopoles are completely eliminated. 
Thus, we introduce continuous variables $\Tilde{a}_{*c}$ on dual links to act as Lagrange multipliers that effect the vanishing of the magnetic current, $(\diff m)_c=0$.
The result is a so-called modified Villain action \cite{Sulejmanpasic:2019ytl,Gorantla:2021svj}:
\begin{align}
    S_{\text{mV}} = \frac{\beta}{2} \sum_p |(\diff a)_p + 2 \pi m_p|^2 - i \sum_p m_p (\diff \Tilde{a})_{*p}
    \label{MV-m}
\end{align}

We now wish to determine the phase of this system for small $\beta$, where center vortices are certain to proliferate. 
We can answer this question immediately by performing a Poisson resummation over the $m_p$ to obtain a dual description of the system:
\begin{align}
    \Tilde{S}_{\text{mV}} = \frac{1}{8\pi^2\beta} \sum_p |(\diff \Tilde{a})_{*p} + 2 \pi \Tilde{m}_{*p}|^2 + i \sum_p (\diff a)_p \Tilde{m}_{*p}
\end{align}
where the $\Tilde{m}_{*p}$ are new integer variables on dual plaquettes. 
This is again a modified Villain action, but now for a $\U(1)$ gauge theory.
We see that here $a$ serves as a Lagrange multiplier to constrain the dual magnetic current $\diff \Tilde{m}$, but the fact $a$ is quantized in units of $\frac{2\pi}{N}$ implies that the dual magnetic current $\diff \Tilde{m}$ should only vanish mod $N$.
Thus, this dual theory describes a $\U(1)$ gauge field coupled to monopoles of charge $N$. 

Now, in the $\U(1)$ description, $\beta$ corresponds to the gauge coupling, and if it is sufficiently small, we are sure to reach a Coulomb phase.
But as we have already said, small $\beta$ in the $\Z_{{} N}$ description is the regime where center vortices proliferate.
We have thus proven that with the complete elimination of monopoles, a Coulomb phase arises when center vortices become light.

\begin{figure}[tbp] 
   \centering
   \includegraphics[width=4in]{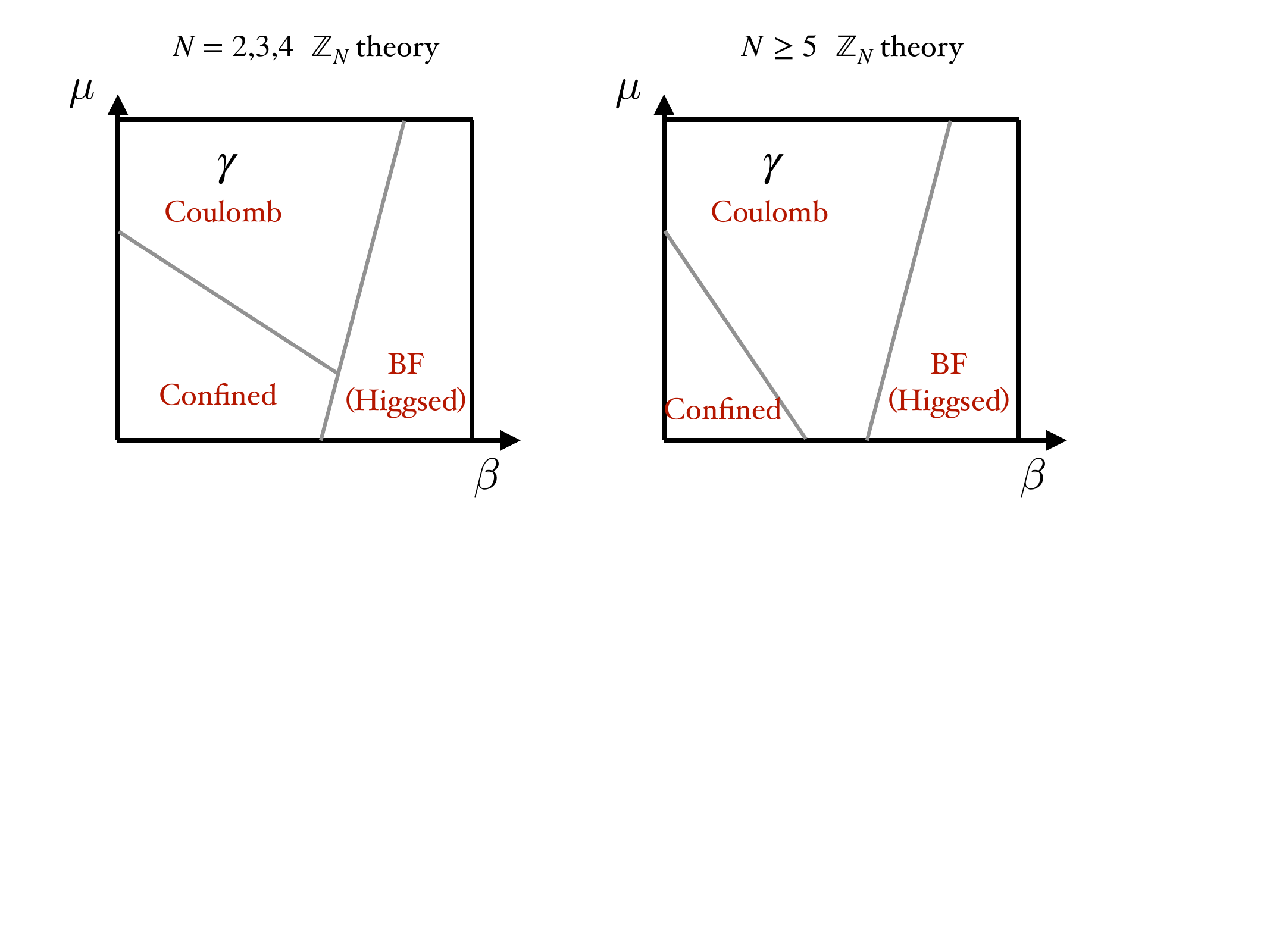} 
   \vspace{-4cm}
   \caption{A cartoon of the phase diagram for 4d $\mathbb Z_N$ lattice gauge theory with a monopole suppression term. }
   \label{fig:phase_diag}
\end{figure}

We now wish to argue that the Coulomb phase should persist if we  allow monopoles into the system but give them a large bare mass, e.g., by adding to $S_{\text{mV}}$ the term
\begin{align}
    \mu \sum_c |(\diff m)_c|^2
\end{align}
To see this, note that when monopoles were absent, the system had a $\tilde{\U}(1)^{[1]}$ symmetry under shifts $\Tilde{a} \to \Tilde{a} + \Tilde{\alpha}$ with $\diff \Tilde{\alpha} = 0 \mod 2 \pi$.
In fact, this symmetry participates in a mixed anomaly with the $\Z_{{} N}$ center symmetry, which already implies that a trivially gapped confined phase is impossible.
In the presence of monopoles, this $\tilde{\U}(1)^{[1]}$ symmetry is explicitly broken but should emerge at low energy if the monopoles are heavy. 
In the dual description, this explicit breaking of $\tilde{\U}(1)^{[1]}$ symmetry should correspond to coupling the gauge field $\Tilde{a}$ to heavy electric matter.
The presence of heavy electric matter in a $\U(1)$ gauge theory cannot change the fact that there is a Coulomb phase for sufficiently weak coupling. 
We therefore conclude that the Coulomb phase persists even when the monopoles of the original $\Z_{{} N}$ description are present. The sketch of the phase diagram is presented in Fig.~\ref{fig:phase_diag}.


One can also ask the question of what deformation of the Wilson-type $\ZZ_N$ action will push the system to the photon phase. A natural candidate for such a deformation would be to identify a monopole current, and then suppress it. Indeed the monopole currents for a $\U(1)$ gauge theory in Wilson formulation have been extensively used in the 80s \cite{DeGrand:1980eq,Barber:1984ak,Grosch:1985cz}, and the definition should carry over to the $\ZZ_N$ lattice gauge theory. It is then reasonable to conjecture that suppressing these monopole currents will lead to a robust photon phase just like in the Villain model. We leave the study of this model for the future.

\section{Conclusions}

Two competing pictures of confinement dating to early days of QCD are the proliferation of monopoles and the proliferation of center vortices.  
Our analysis provides a rigorous result in this controversy. 
Lattice simulations have shown that the removal of center vortices leads to the vanishing of the string tension and loss of confinement \cite{DelDebbio:1996lih, DelDebbio:1998luz,deForcrand:1999our, Engelhardt:1998wu}. 
This numerical fact is often interpreted as support that center vortices are the whole story to confinement (i.e., that they are both necessary and sufficient\footnote{We do note that it was observed in \cite{deForcrand:1999our} that the removal of center vortices also prevents the condensation of monopoles.}), and sometimes as counterevidence against the monopole mechanism.  
Our analysis based on the deformation class of the topological \emph{BF} theory leads to a different conclusion.  
We have shown that the vortex proliferation in the absence of monopole proliferation leads to a phase transition from the topological \emph{BF} (Higgs) phase to the Coulomb phase. 
Despite the fact that center vortices themselves do not drive confinement, they still play a crucial role.  
Their condensation prepares the vacuum in which monopoles can condense. 
If the monopoles do condense on top of the vortex condensate, then the Coulomb potential turns into a linearly confining potential at large distances. 
Otherwise, the Coulomb phase prevails.




We can easily generalize our discussion to include the $\theta$ angle. 
Then, the monopole condensation will be replaced by the dyon condensation. 
We can also generalize our construction to QFTs with both 0-form and  $\Z_{{} N}$ 1-form global symmetry. 

It would be interesting to check our picture numerically. 
It is well established by now that, for sufficiently large $N$, a photon phase emerges in the $\Z_{{} N}$ gauge theory for some intermediate coupling $\beta$ \cite{Elitzur:1979uv,Creutz:1979zg,Horn:1979fy,Frohlich:1982gf}. 
It would be interesting to see that this phase is associated with monopole non-proliferation. 
Further, it would be useful to check the existence of the photon phase for small $N$, proposed in Fig.~\ref{fig:phase_diag}. 
Finally, one could devise checks in the $\SU(N)$ gauge theory. 
Of course, here it is tricky to define precisely what is meant by either center vortices of monopoles\footnote{This is a bit like asking what domain walls are in the phase where a discrete symmetry is not spontaneously broken.} \cite{Chernodub:1997ay}. 
This is not surprising; they are well defined objects only in phases where they are heavy, and where they have no chance of proliferating in the vacuum. 
We suspect that the qualitative picture will not depend on the precise definition of these objects. 
So one could pick an abelian projection, and suppress monopoles to show that a photon phase emerges, provided that this suppression does not somehow cause the center vortices to be suppressed as well.

\acknowledgments

We would like to thank Aleksey Cherman, Massimo D'Elia, Philippe de Forcrand, Theo Jacobson, Biagio Lucini and Nathan Seiberg for their comments and to Jeff Giansiracusa, David Lanners and Tyler Helmuth for discussions. 

M.N. thanks Durham University for their hospitality where much of this work was completed. 
T.S. is supported by the Royal Society University Research Fellowship and in part by the STFC consolidated grant ST/T000708/1.
M.\"U. is supported by U.S. Department of Energy, Office of Science, Office of Nuclear Physics under Award Number DE-FG02-03ER41260. 

\appendix

\section{Basics of \emph{BF} theory}

\label{sec:BF_review}

This appendix reviews the quantization of \emph{BF} theory. 
Let us first establish the basic correlation function of the theory:
\begin{align}
    \langle \exp( i \textstyle{\int_V} b ) \exp ( i \textstyle{\int_C} a )\rangle 
    = \exp \{ i \frac{2\pi}{N} \Link (V ,C) \}
    \label{CF}
\end{align}
where $C$ is a 1-cycle and $V$ is a $(d-2)$-cycle.
For simplicity, let us take $\RR^d$ for Euclidean spacetime.
Then $C$ and $V$ are both boundaries, and the expectation values of both $\exp(i \int_C a)$ and $\exp(i\int_V b)$ will be assumed to be unity. 
Now pick a 2-chain $D$ with boundary $C$ so that we can write $\int_C a = \int_D \diff a$.
Then by shifting $b \to b + \frac{2\pi}{N} \delta[D]$ where $\delta[D]$ is Poincar\'{e} dual to $D$, we can cancel the $\exp(i \int_C a)$ insertion, bringing the correlation function into the form
\begin{align}
    \exp(i \tfrac{2\pi}{N} \textstyle{\int_V} \delta[D]) \ \langle \exp(i\textstyle{\int_V} b) \rangle
\end{align}
Since as we have said, the expectation value of $\exp(i\int_V b)$ is unity, the only nontrivial thing is the first factor. 
All that remains is to recognize that
\begin{align}
    \int_V \delta[D] = \int \delta[V] \wedge \delta[D]
\end{align}
is nothing but the linking number $\Link(V,C)$ of $V$ and $C$.

It is useful also to discuss the theory in the Hamiltonian framework.
We formulate the theory in real time on a spatial manifold $X_{d-1}$ which we assume to be compact and, for simplicity, torsion-free.
Choose 1-cycles $C_i$ giving a basis of $H_1(X,\ZZ)$, and let $V_i$ be the dual $(d-2)$-cycles so that $\int_X \delta[C_i] \wedge \delta[V_j] = \delta_{ij}$.
[The $V_i$ thus give a basis of $H_{d-2}(X,\ZZ)$.]
Since the equations of motion imply that $a$ and $b$ must be flat, they are completely specified by their holonomies $\alpha_i \equiv \int_{C_i} a,\ \beta_i \equiv \int_{V_i} b$.
Note that $\alpha_i,\beta_i$ shift by integer multiples of $2\pi$ under large gauge transformations, so they must be viewed as angular variables. 
Expressing the \emph{BF} action (in Lorentzian signature) in terms of the $\alpha_i, \beta_i$ gives
\begin{align}
    S 
    &= \frac{N}{2\pi} \sum_i \int dt \ \beta_i \frac{d \alpha_i}{dt}
\end{align}
The canonical commutation relations are therefore given by $[\alpha_i , \beta_j] = i \frac{2\pi}{N} \delta_{ij}$.
Upon exponentiation these give the so-called clock-and-shift relations
\begin{align}
    &\exp(i \alpha_i) \exp(i\beta_j) = \exp(i\beta_j) \exp(i\alpha_i) \exp (-i\tfrac{2\pi}{N}\delta_{ij})
\end{align}
We construct the Hilbert space by postulating a unique state $|\{p_i\}\rangle$ for each set of simultaneous eigenvalues $\exp(i \frac{2\pi}{N}p_i)$ of the operators $\exp(i\alpha_i)$.
The action of $\exp(i\beta_i)$ on these states shifts $p_i$ by $-\delta_{ij}$. 

\section{More about confinement in two dimensions}
\label{sec:2dconf}
In this appendix, we give an explicit computation of the Wilson loop average in the 2d deformed \emph{BF} theory.
This will elucidate the interpretation of the system as a dilute gas of center vortices as well as provide a very general perspective on 2d confinement.  
We will essentially be presenting a special case of the analysis given in Ref.~\cite{Cherman:2021nox}, which studied the effect of adding topological pointlike operators to the action.

We shall adopt the following notation:
\begin{align}
    V(x) &\equiv e^{ib(x)} \\
    W(C) &\equiv e^{i\int_C a} \\
    \omega &\equiv e^{2\pi i/N} \\
    \delta_N(n) &\equiv \sum_{k \in \Z} \delta_{n,Nk}
\end{align}
In fact, we can incorporate the presence of a $\theta$ parameter by taking the Lagrangian
\begin{align}
    \mathcal{L} = \frac{iN}{2\pi}\, b \, \diff a - \zeta (e^{i\theta/N} V + e^{-i\theta/N} V^*)
\end{align}

As a first step, we must evaluate the partition function. 
Using the selection rule $\langle V^p \rangle_* = \delta_N(p)$, where the subscript $*$ denotes quantities evaluated in the undeformed \emph{BF} theory, we write
\begin{equation}
    Z = Z_* \sum_{n,\bar{n} = 0}^{\infty} \frac{1}{n!\bar{n}!} \zeta^{n + \bar{n}} e^{i(n - \bar{n})\theta/N} {\cal A}_M^{n + \bar{n}} \delta_N(n - \bar{n})
\end{equation}
where $\mathcal{A}_M$ is the area of the space time manifold $M$. 
If we replace $\delta_N(n-\bar{n})$ by its discrete Fourier expansion $N^{-1}\sum_{k = 0}^{N-1} \omega^{(n-\bar{n}) k}$, the sum over $n,\bar{n}$ becomes two independent exponential series, and we find
\begin{equation}
    Z = Z_* N^{-1}\sum_{k=0}^{N-1} \exp (- \varepsilon_k {\cal A}_M),
\end{equation}
where 
\begin{equation}
    \varepsilon_k(\theta) \equiv -2 \zeta \cos \biggl( \frac{\theta+ 2 \pi k}{N} \biggr)
    \label{branches}
\end{equation}
are the energy densities of the $N$ energy eigenstates $|k\rangle$. 

We turn now to the Wilson loop expectation value:
\begin{multline}
    \langle W^q(C) \rangle = \frac{Z_*}{Z} \sum_{n,\bar{n} = 0}^{\infty} \frac{1}{n!\bar{n}!} \zeta^{n + \bar{n}} e^{i(n - \bar{n})\theta/N} \\
    \times \int \prod_{i=1}^n d^2x_i \prod_{j=1}^{\bar{n}} d^2y_j \langle W^q(C) \prod_{i=1}^n V(x_i) \prod_{j=1}^{\bar{n}} \bar{V}(y_j) \rangle_*
\end{multline}
Let us take $C$ to be the boundary of a disk $D$ of area ${\cal A}$ and set ${\cal A}' = {\cal A}_M-{\cal A}$. 
To evaluate the correlator, note that by Eq.\eqref{CF} we can replace each $x_i\in D$ or $y_j \in D$ with any $x_i'\notin D$ or $y_j' \notin D$ at the cost of picking up a factor $\omega^q$ or $\omega^{-q}$. 
Having done this for every $x_i,y_j \in D$, we find
\begin{multline}
    \langle W^q(C) \prod_{i=1}^n V(x_i) \prod_{j=1}^{\bar{n}} \bar{V}(y_j) \rangle_* \\ 
    =
    \prod_{i=1}^n (\omega^q)^{\chi(x_i)} 
    \prod_{j=1}^{\bar{n}} (\omega^{-q})^{\chi(y_j)}
    \delta_N(n-\bar{n})
\end{multline}
where $\chi(z)$ is 1 for $z \in D$ and 0 otherwise. 
The integrals over the $x_i,y_j$ are now decoupled and evaluate to
\begin{multline}
    \int \prod_{i=1}^n d^2x_i \prod_{j=1}^{\bar{n}} d^2y_j \langle W^q(C) \prod_{i=1}^n V(x_i) \prod_{j=1}^{\bar{n}} \bar{V}(y_j) \rangle_* \\
    = (\omega^q {\cal A} + {\cal A}')^n (\omega^{-q} {\cal A} +{\cal A}')^{\bar{n}} \delta_N(n - \bar{n})
\end{multline}
From this, we find
\begin{equation}
    \langle W^q(C) \rangle
    = \frac{\sum_{k = 0}^{N-1}
    \exp ( - \varepsilon_{k+q} \mathcal{A} - \varepsilon_{k} \mathcal{A}') }{\sum_{k=0}^{N-1} \exp (-\varepsilon_{k} \mathcal{A}_M)}
    \label{eq:WilsonLoopAverage}
\end{equation}
Let  $k_0 $ denote the ground state for some range of $\theta$ angle according to \eqref{branches}.
Then,  taking the limit $\mathcal{A}_M \to \infty$ while keeping $\mathcal{A}$ fixed, we calculate the expectation value of the Wilson loop to be
\begin{equation}
    \langle W^q(C) \rangle = \exp \{- (\varepsilon_{k_0+q} - \varepsilon_{k_0}) {\cal A} \}
    \label{eq:AreaLaw}
\end{equation}
  which is the area law of confinement. 

We emphasize that  in 2d, both center vortices and Wilson loops have special properties, which do not extend to $d \geq 3$. 
In  particular, in 2d, the Wilson line may be interpreted as a (nondynamical) domain wall defect between the state $|k_0\rangle$ outside and the state $|k_0 + q\rangle$ inside the loop. 
Furthermore, center vortices are pointlike objects, instantons, which can proliferate regardless of coupling. 
These two make the above demonstration of confinement a manifestly 2d mechanism. 
In fact, in the discussion of $\Z_{{} N}$ lattice gauge theories in $d=3, 4$, center vortices are extended objects, and Wilson lines no longer serve as domain walls between vacua. 
In these cases, we show that proliferation of vortices without proliferation of monopoles leads to a Coulomb potential rather than a linearly growing potential.

\section{3d \boldmath{\texorpdfstring{$\Z_{{} N}$}{ZN}} lattice gauge theory with or without monopoles}

We would like to use our   findings in the context of the deformation
of \emph{BF}  theory to obtain a deeper  analytical understanding of
numerical simulations of $3$d  $\Z_{{} N}$ lattice gauge theory
\cite{Bhanot:1980pc}.
It is known that the theory has two phases, for $\beta>\beta_c$, it is in the
deconfined  phase, and for $\beta<\beta_c$, it  is in the
confined phase \cite{Bhanot:1980pc}.  It is also known that both monopoles and
center-vortices must play some role in its non-perturbative dynamics, see e.g.
\cite{Chernodub:2005be, Athenodorou:2023jrb}.
However, their precise role is  still unclear. Our method will help
us to identify  their  effects  in dynamics unambiguously. 


A heuristic way to see the existence of the phase transition is tied
with the center vortices is as follow:  The action of a minimal   vortex loop  of
length $L$ is  $S= \beta (1- \cos \frac{2 \pi}{N} )L$. 
The number of non-backtracking loops of length $L$  passing through a fixed point is approximately $(2d-1)^L$ where $2d$ is the number of nearest neighbors, hence the entropy associated with it is  approximately  $e^{ \log(2d-1) L} = e^{ \log(5)
L}  $. 
Therefore, the single vortex contribution to the partition
function function is of the form
  \begin{align}
 Z_{\text{one vortex}} \sim   e^{ \log (5)  L}   e^{  - \beta (1-
\cos \frac{2 \pi}{N})L }.
 \label{PF}
  \end{align}
 The action (energy times length) aims to suppress the proliferation
of vortices, while the entropic factor enhances it. This is the
standard Peierls instability argument, leading to a phase transition
at the scale
 \begin{align}
 \beta_c =  \frac{  \log 5}{ 1- \cos \frac{2 \pi}{N} }
 \end{align}
 Indeed, this heuristic result agrees well with lattice data
\cite{Bhanot:1980pc}.
  At this stage, one can say that vortices play a crucial role in the
phase transition.  The fact that the vortices proliferate in the
confined phase and their  non-trivial mutual statistics with the
Wilson loops   $\exp \{i \frac{2\pi}{N} \Link (V,C)\}$
  led many authors to speculate that the proliferation of the center
vortices  is the  mechanism of confinement \cite{Greensite:2003bk,Engelhardt:1998wu}. 
  On the other hand,  in  the $N \rightarrow \infty$ limit, $\Z_{{} N}$
theory becomes   $\U(1)$ gauge theory, and 
 the phase transition disappears.
   The $\U(1)$   theory is   confining at any value of the coupling
constant $\beta$ and this occurs via the  proliferration of the
monopoles \cite{Polyakov:1975rs}.  In fact, in the solution of the $\U(1)$ model, the center
vortices do not even make an appearance.
     This type of situation led to many controversies regarding the
roles of topological defects in gauge theories over the years. For a
review, see \cite{Greensite:2003bk}.   Can we say something   {\it
precise}  on the roles of monopoles and center-vortices?

Similar to 4d discussion, we wish to isolate monopoles from the vortices. Let us first eliminate monopoles completely. To do so, 
we introduce continuous variables $\Tilde{\sigma}_{*c}$ on dual sites  as Lagrange multipliers. 
This can be used to  enforce the vanishing of the magnetic flux through cubes  on the lattice, i.e., eliminate monopoles, as in \eqref{MV-m}.   
Again,  by  performing a Poisson resummation over the $m_p$, we obtain 
\begin{align}
    \Tilde{S}_{\text{mV}} = \frac{1}{8\pi^2\beta} \sum_p |(\diff\Tilde{\sigma})_{*p} + 2 \pi \Tilde{m}_{*p}|^2 + i \sum_p (\diff a)_p \Tilde{m}_{*p}
\end{align}
where  $\Tilde{m}_{*p}$ are new integer variables on dual links. 
This is  abelian duality on the lattice. We started with 3d $\Z_{{} N}$ lattice gauge theory and mapped it to an XY-model. In the dual formulation, the original $\Z_{{} N}$ valued $a$ appears  as a Lagrange multiplier. The curvature of $\diff m$ of the integers $m$ is associated with compact-scalar vortices \cite{Gross:1990ub,Sulejmanpasic:2019ytl}. Since it is $\Z_{{} N}$ valued, it essentially projects out compact-scalar-vortices of charges which are not multiple of $N$. Therefore, the  dual theory describes XY-model 
 coupled to  charge $N$ global vortices. The duality also swaps $\beta \leftrightarrow  1/(4\pi^2\beta)$.

\begin{figure}[tbp] 
   \centering
   \includegraphics[width=4in]{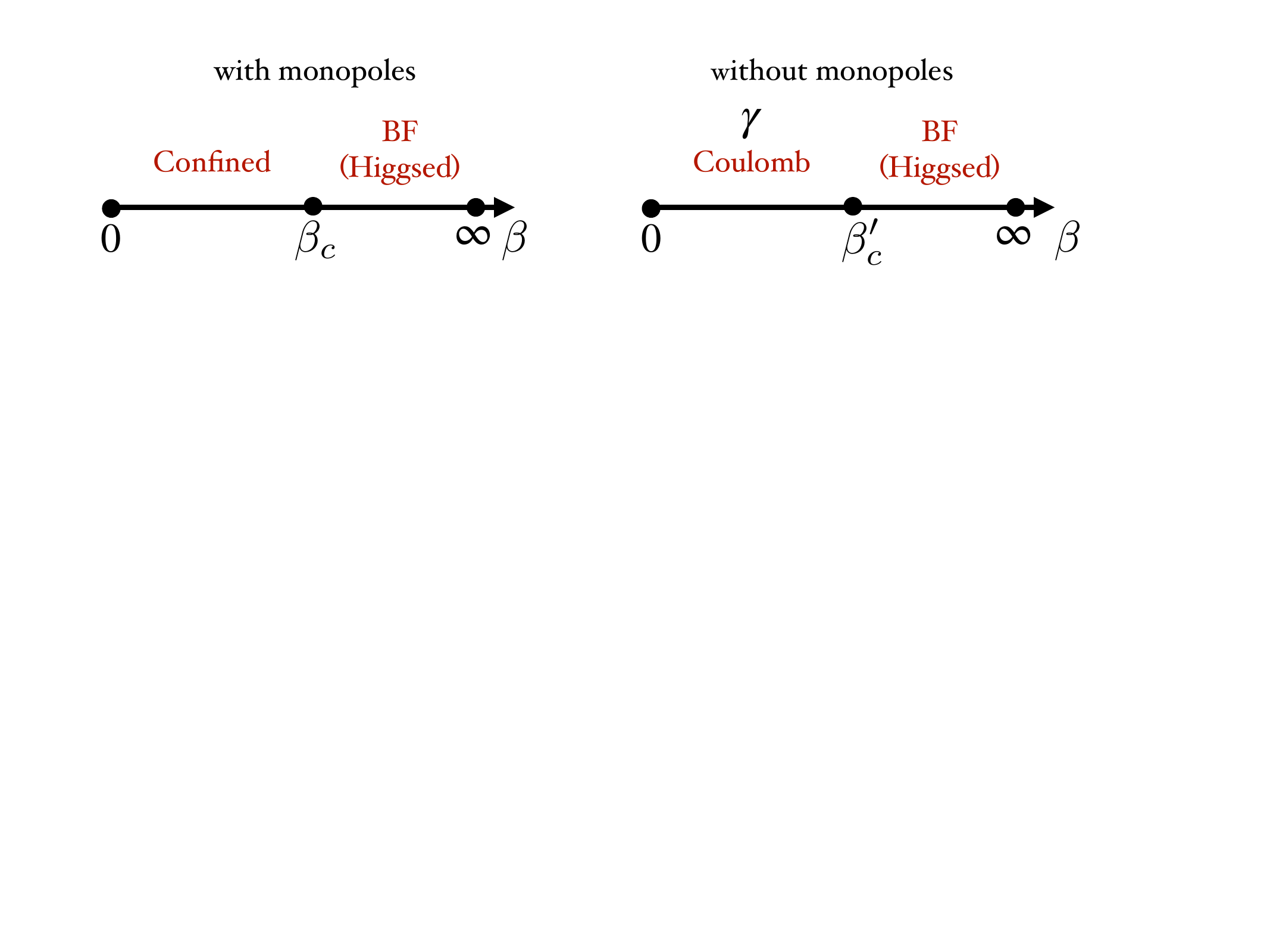} 
   \vspace{-5.5cm}
   \caption{Phases of 3d  $\mathbb Z_N$ lattice gauge theory with and without monopoles. }
   \label{fig:phases3d}
\end{figure}

In the original $\Z_{{} N}$ gauge theory, small $\beta$ is the strongly coupled regime where center vortices proliferate.  
In the XY-model description, small $\beta$  corresponds to a weak coupling regime where vortices are suppressed, and the XY-model  is gapless. On the other hand for sufficiently large $\beta$, the model is in the topological \emph{BF} phase, so that the minimal scenario is one with two phases (See Fig.~\ref{fig:phases3d}.)



 In the absence of monopoles,   the statistical field theory at
our disposal is a theory of vortex loops. 
These vortex loops have non-trivial mutual statistics with the Wilson
loops.  
  In the topological \emph{BF} (Higgs) phase,
 the interaction between the external test charges is constant, $V(r)=
V_0$ and there is no force between them. 
 However, once the vortices proliferate,  this
turns the constant Higgs potential into the Coulomb potential,
$V(r) \sim  \log (r)$  (and not to linearly rising potential.)   
Our result at this point disagrees with standard lore in center-vortex literature  
 in $d=3,4$ dimensions \cite{Greensite:2003bk,Engelhardt:1998wu}. Our analysis shows that in the absence of monopoles, 
 the proliferation of vortices does not generate linear confinement.

Now, we can incorporate monopoles  by adding monopole operators  to the Lagrangian:
\begin{align}
\Delta {\cal L} =  -\zeta e^{i\Tilde{\sigma}} + {\rm c.c.}
 \end{align}
 where $\zeta$ is monopole fugacity.  This operators explicitly breaks $\tilde{\U}(1)$
topological shift  symmetry $\Tilde{\sigma} \rightarrow\Tilde{\sigma} + \epsilon$.  For $\beta < \beta_c$, proliferation of monopoles 
generate  linear confinement $V(r) \sim  r$  and  mass gap for the dual photon.  For  $\beta > \beta_c$, however, in the topological Higgs phase,  monopoles are confined into neutral clusters held together by heavy center vortices. 

Two main results  are following.
\begin{itemize}
\item[a)] The condensation of the  center-vortices  leads to a phase transition from 
the Higgs phase to Coulomb phase (in the absence of monopoles)
and from the Higgs phase to the confining phase (in the presence of  monopoles).  
\item[b)]   Monopoles can  proliferate if and only if the
center-vortex  condenses.  In the phase 
where center-vortices are heavy, monopoles are confined into neutral clusters.  Condensation of the vortices  activates  the monopoles, and their proliferation leads to confinement. 
\end{itemize}
It is remarkable that monopoles lead to both confinement and mass gap
in the 3d $\Z_{{} N}$ lattice gauge theory. But this happens  if and only if the center vortices is already  condensed.

The reason why we do not talk about the 
center-vortices in the Polyakov model is because the theory is already in the phase where the center-vortex is condensed. If we were to consider a
 symmetry respecting deformation of Polyakov model with two adjoint Higgs scalars which breaks $\SU(N) \rightarrow \Z_{{} N}$, we would end up in a topological Higgs phase. Gradually removing one of scalars to reduce the theory to the standard Polyakov model, we would end up with   $\SU(N) \rightarrow \U(1)^{N-1}$. In this process, the center-vortices must become light, leading to a phase transition and at the same time, activating the monopoles. 

\section{The center vortices and the Wilson loops}\label{app:center_vortex_model}

Here we make precise how center vortices interact with Wilson loops. Namely the standard argument goes as follows. Consider an ensemble of vortex-type configurations: vortex-instantons in 2d, vortex-particle in 3d and vortex-string in 4d. To model this on the lattice, take an integer variable $k_p$ defined on a plaquettes of the lattice labeled by $p$, i.e. it is a 2-cochain. In dimensions $d\neq 2$ we impose the constraint that $(\diff k)_c=0$, so that this integer field corresponds to a closed loop in 3d (particle worldline) or  closed surface  in 4d (string worldsheet). Note that the vortex itself naturally lies on the dual lattice. The partition function of this ensemble is given by
\begin{equation}
Z=\sum_{\{k_p\}} e^{-\frac{\beta}{2} |k_p|^2}\;,
\end{equation}
where we picked that the weight is Gaussian. Since we will be interested in the limit $\beta\rightarrow 0$, where vortices are not suppressed and proliferate, we do not expect that the exact shape of the weight matters. The sum over $k_p$ implicitly has a constraint $(\diff k)_c=0$ in $3d$ and $4d$.
Now we associate with the system a probe object: the Wilson loop, which can be thought of as a closed loop on the lattice. The Wilson loop has to have topological correlation functions with the vortex. To encode this, let $\Sigma$ be a surface built up out of the plaquettes of the lattice, whose boundary $\partial \Sigma$ is the contour of the Wilson loop. To this $\Sigma$ surface we associate an integer\footnote{The $\star$ operator is a lattice analogue of the Hodge star and it associates the plaquette $p$ of the lattice with the $(d-2)$-cell of the dual lattice, i.e. a link in $3d$ and a plaquette in $4d$.} $C_{\star p}$, which is $0$ if $p \notin \Sigma$ and $n$ if $p\in \Sigma$, where $n$ is the charge of the Wilson loop.  The defining property of the Wilson loop is that it has a topological correlation function with the center vortex producing a $e^{i\frac{2\pi n}{N}}$ phase whenever it links with the center vortex, i.e. the Wilson loop expectation value is
\begin{equation}
\avg{W_n}=\frac{1}{Z}\sum_{\{k_p \}}e^{-\frac{\beta}{2}|k_p|^2+i\frac{2\pi }{N}\sum_{p}C_{\star p}k_p}\;.
\end{equation}
Now consider the $2d$ case first. In this case $k_p$ is unconstrained and we can perform the Poisson resummation
\begin{equation}
\sum_{k_p} e^{-\frac{\beta}{2}|k_p|^2+i\frac{2\pi}{N}C_{\star p} k_p}=\frac{1}{\sqrt{2\pi \beta}}\sum_{q_{\star p}}e^{-\frac{1}{2\beta}\left(q_{\star p}+\frac{C_{\star p}}{N}\right)^2}
\end{equation}
where we introduced the Poisson dual integers $q_{\star p}$ on the sites $\tilde x=\star p$ of the dual lattice. In the limit that $\beta$ is small\footnote{In $2d$ one can perform an exact calculation, with the same conclusion, but this limit is simplifying.}, the sum over $q_{\tilde x}$ is dominated by $q_{\tilde x}=0$, as long as\footnote{If the charge of the Wilson loop is not in this range, then we can always redefine $q$ to bring it to this range, so that the result depends only on $k\bmod N$.} $-N/2<n<N/2$. The Wilson loop expectation value is then given by
\begin{equation}
\avg{W_n}\sim e^{-\frac{n^2 A}{2N^2\beta}}\;,
\end{equation}
where $A$ is the area of $\Sigma$. This is the famous area law.

Now if we try to perform the same calculation in higher dimensions, we must impose the constraint $(\diff k)_c=0$. To do this we introduce the field $\tilde a_{\star c}$ on the dual cells of the cubes $c$ of the lattice. In other words these are dual sites $\tilde x$ in $3d$ and dual links $\tilde l$ in $d=4$. The constraint is imposed as follows
\begin{equation}
\int_0^{2\pi} \frac{\diff \tilde a_{\star c}}{2\pi}e^{i\tilde a_{\star c}(\diff k)_{c}}\;.
\end{equation}
Now when we do the Poisson resummation, the partition function with the Wilson loop insertion is given by
\begin{equation}
\sum_{\{q\}}e^{-\frac{1}{2(2\pi)^2\beta}\sum_p|(\diff \tilde a)_{\star p}+2\pi q_{\star p}+\frac{2\pi}{N}C_{\star p}|^2}\;.
\end{equation}
In 4d the model we get is in fact an $\U(1)$ abelian gauge theory Villain model, while in $3d$ it is the compact scalar model. In the limit where $\beta\rightarrow 0$, and without the Wilson loop (i.e. $C=0$) the model has a suppression of the Villain variables $q$. Since Villain variables are associated with monopoles and vortices \cite{Cardy:1976ni,Cardy:1981qy,Gross:1990ub,Sulejmanpasic:2019ytl}, in this regime we get a $\U(1)$ gauge theory or a compact scalar theory on the lattice with suppressed monopoles/vortices, which is famously gapless. This is the first indication that center vortices are not enough for confinement. Further, the Wilson loops are equivalent to the insertion of $C_{\star p}$ integers, which are such that $(\diff C)_{\star l}=\sum_{l'\in \partial \Sigma}\delta_{l,l'}$. This is equivalent to the insertion of the 't Hooft loop in the $\U(1)$ lattice model, which in the gapless $\U(1)$ phase has a Coulomb perimeter law.


\section{Exotic phases}\label{app:photons}

 \subsection{Multiple photons and partially broken \boldmath{\texorpdfstring{$\Z_{{} N}$}{ZN}} 1-form symmetry}

In the main text we discussed the most robust and generic phases. However more exotic scenarios are possible. Here we discuss some of these possibilities. 

 So far we discussed only the condensation of charge $1$ vortices. However 1-form symmetry does not prohibit the existence of higher-charge objects. A charge $q$ center vortex is equivalent to the emergence of a gauge field $a^{(q)}$. Only charge $q\bmod N$ is meaningful\footnote{By this we mean that the higher charges of center vortices are equivalent $\bmod N$. If we did introduce such higher charge extended objects, only a linear combination of them would couple to $b$. While such extended objects would induce extra photons, they do not naturally couple to the rest of the system.} and the effective action is given by
\begin{multline}
    S = \frac{i}{2\pi} \int b\,(\diff a^{(1)}+2\diff a^{(2)}+\cdots (N-1)\diff a^{(N-1)}+N\diff a)\\
    +S_0[\diff a^{(1)},\diff a^{(2)},\cdots,\diff a^{(N-1)},\diff a]
\end{multline}
 the action $S_0$ is necessarily higher order in derivatives. The above is not quite right, because in reality the gauge fields $a^{(q)}, q=1,\cdots, N-1$ are actually coupled to matter fields. If this wasn't the case, we would have an emergent $\ZZ_{q}^{[1]}$ symmetry associated with every $a^{(q)}$. The statement of charge $q$ center-vortex condensation is equivalent to the statement that these matter fields are heavy and the above effective action is correct. We now discuss four special cases.
 \begin{enumerate}
 \item All charges condense
 \item Only charge $q>1$ condenses, with $\GCD(q,N)=1$
 \item Only charge $q>1$ condenses, with $\GCD(q,N)\ne 1$.
 \end{enumerate}
 In the case 1. if all charges condense we we have $N$, \emph{BF} terms. Integrating over $b$ imposes a constraint that $\diff a^{(1)}+\diff a^{(2)}+\cdots (N-1)\diff a^{(N-1)}+N\diff a=0$, which effectively eliminates $1$ photon field, and leaves $N-1$ photons in the theory. The effective theory is that of $N-1$ photon fields! However, one should also discuss what happens to the monopoles of these photons, which can further gap them. We will not discuss this scenario here in detail. Instead we will construct the lattice model where $N-1$ photons do emerge. 
 
A special scenario of case 2 already appeared before in the main text, where $q=1$ and results in a $\U(1)$ gauge theory, which either may be confining (monopoles proliferate) phase or the photon phase (monopoles do not proliferate). Indeed, integrating out the $b$-field introduces a single constraint $q\diff a^{(q)}+N\diff a=0$, allowing one gauge field to emerge. Importantly the \emph{BF} theory was rendered trivial. Here we can see this by noting that the combination $c=qa^{(q)}+Na$ is a new $\U(1)$ gauge field, with standard quantization if $\GCD(q,N)=1$. The \emph{BF} term is then $i\int dc\wedge b$, which is a trivial \emph{BF} theory. This is not the case when $G=\GCD(q,N)\ne 1$. Indeed in this case it is the field $c=\frac{q}{G}a^{(q)}+\frac{N}{G}a$ is a properly quantized gauge field, and the remaining \emph{BF} term is $iG
\int dc\wedge b$, indicating that $\ZZ_{{} N}^{[1]}$ breaks to $\ZZ_{{} {N/G}}^{[1]}$.

\subsection{\boldmath{\texorpdfstring{$N-1$}{N-1}} photons in a lattice model}

Now let us discuss a lattice model that has multiple emergent photons. 
To motivate the model, note that the condensation of charge $q$ center vortices in continuum is in some sense describing the bound state of center vortices. 
To facilitate interactions of center vortices, we will introduce, instead of just the usual Wilson term $\beta\{1-\cos (\diff a)_p\}$, $N-1$ terms $\beta_r\{1-\cos(r \diff a)_p\}$, where $r=1,\cdots, N-1$.
Thus, consider the action
\begin{equation}
S=\sum_{r=1}^{N-1} \sum_p \beta_r\Big[1-\cos(r \diff a)_p\Big]\;.
\end{equation}
The action clearly associates a different weight for different charge center vortices. A charge $q$ center vortex has the action $S_q= v\sum_r\beta_r(1-\cos(\frac{2\pi qr}{N}))$ where $v$ is the world-volume of the center vortex. It is clear that by dialing $N-1$ parameters $\beta_r$, we can always tune $S_q$ for any $q$.

Now let us replace this action with a Villain type action, expected to be in the same universality class. This is given by
\begin{equation}
S=\sum_{r=1}^{N-1}\sum_p\frac{\beta_r}{2}(r \diff a_p-2\pi m^r_p)^2\;,
\end{equation}
where $m^{r}_p$ are integers on plaquattes. 
Each of these integer variables can be associated with a monopole. 
Imposing the constraints $(\diff m^{r})_c=0$ via Lagrange multipliers, and Poisson re-summing, we arrive at the dual action
\begin{multline}
\Tilde{S} =\sum_{r=1}^{N-1} \frac{1}{8\pi^2\beta_r}\sum_{\tilde p}( \diff \tilde a^r_{\tilde p}-2\pi \tilde m^r_{\tilde p})^2\\+i\sum_{r=1}^{N-1}\sum_{\tilde p}(\diff a^{(r)})_{\star \tilde p}r\tilde m^{r}_{\tilde p}\;,
\end{multline}
The action amounts to the action of $N-1$ Villain $\U(1)$ gauge theories, which will be in the photon phase if $\beta_{q}$ is sufficiently small. The phase cannot change if we now introduce back the heavy monopoles. 

\bibliography{refs.bib}

\end{document}